\documentclass{ws-ijmpe}

\newcommand{\kF}{k_{F}}                                % fermi momentum
\newcommand{\sclgth}{a}                                % scattering length
\newcommand{\reff}{r_{\textrm{ef\mbox{}f}}}              % effective range
\newcommand{\fm}{\,\textrm{fm}}                          % fermi unit
\renewcommand{\vec}[1]{\textbf{#1}}                    % Vectors
\newcommand{\vecp}[1]{\vec{#1}^{\prime}}               % Vector prime
\newcommand{\latconst}{b}                              % lattice const
\newcommand{\Ns}{N_{s}}                                % lattice dim
\newcommand{\dens}[2]{\hat{n}_{#2}(#1)}                % particle occupation operator
\newcommand{\eqn}[1]{Eq.~(\ref{#1})}                   % equation abbreviation
\newcommand{\rrel}{\vec{R}}                            % realitve position
\newcommand{\Tr}{\mathrm{Tr}}                          % Trace
\newcommand{\pcut}{p_{\textrm{cut}}}                     % momentum cut-off

\begin{document}

\markboth{G. Wlaz{\l}owski, P. Magierski}{Effective range expansion for the interaction defined on the lattice}

\catchline{}{}{}{}{}
%%%%%%%%%%%%%%%%%%%%%%%%%%%%%%%%%%%%%%%%%%%%%%%%%%%%%%%%%%%%%%%%%%%%

\title{EFFECTIVE RANGE EXPANSION FOR THE INTERACTION DEFINED ON THE LATTICE}
\author{\footnotesize GABRIEL WLAZ{\L}OWSKI and PIOTR MAGIERSKI}

\address{Faculty of Physics, Warsaw University of Technology, ul. Koszykowa 75\\
00-662 Warsaw,
Poland\\
e-mail: gabrielw@if.pw.edu.pl, magiersk@if.pw.edu.pl}

\maketitle

\begin{history}
\received{(received date)}
\revised{(revised date)}
%\accepted{(Day Month Year)}
%\comby{(xxxxxxxxxx)}
\end{history}

\begin{abstract}
The relation between the interaction parameters for fermions on 
the spatial lattice and the two-body $T$ matrix is discussed.
The presented method allows determination of the 
interaction parameters through  
the relatively simple computational scheme which
include the effect of finite lattice spacing. 
In particular the relation between the interaction parameters and 
the effective range expansion parameters is derived in the limit of large lattices.
\end{abstract}

\section{Introduction}
One of the most promising approaches to describe the nonrelativistic many-fermion system is to define the problem on the spatial lattice and subsequently apply the Monte Carlo technique to probe the configuration space.
This approach is particularly efficient when dealing with strongly interacting and dilute systems. In the
limit of low densities the interaction between particles has predominantly two-body character. 
Moreover it can be
described by few parameters only, characterizing the low energy physics of two-body collisions. 
For example in the case of trapped fermionic atoms, which have been recently the subject of intensive 
theoretical studies (see \cite{gps} and references therein), the scattering length is the only parameter which determines the interatomic interaction. 
Similarly, the physics of dilute neutron matter at densities corresponding to 
$ \kF\lesssim 0.6\fm^{-1}$ is captured by two parameters: scattering length and effective range in the $^{1}S_{0}$ channel. 

The basic problem of the lattice formulation is 
the determination of the Hamiltonian parameters in order to describe properly
the scattering phase shifts of two-body collisions. The prescription given by L\"{u}scher\cite{Luscher} 
relates the discrete energy spectrum of two-particle states in a box 
to the scattering phase shifts. However
in order to use the L\"{u}scher's formula, the exact two-particle energy spectrum
has to be determined first, which in general is a nontrivial task. 
Another prescription, given by Seki and Kolck for the case of large lattices, is more 
suitable for the low-energy nuclear problems.\cite{SekiKolck} In this approach the interaction parameters are determined by consistently applying the effective field theory power counting rules.

In this paper we present an alternative approach to determine the two-body interaction parameters. 
The method is especially well suited to deal with the discrete form of the interaction, commonly used 
in the lattice calculations.\cite{MullerEtAl,AbeSeki,WlazlowskiMagierski} All \textit{finite lattice spacing} effects are properly included in the limit of large lattice volume. The final prescription bypass the tedious L\"{u}scher's algebra leading to a simple expression convenient for practical applications.

\section{Interaction on the lattice}
To capture the low energy physics of a Fermi system let us consider the interaction which depends only 
on the relative coordinates of two particles:
\begin{equation}
 \hat{V}=\frac{1}{2}\sum_{\lambda,\lambda^{\prime}=\uparrow\downarrow}\int  d^{3}\vec{r}d^{3}\vecp{r}\,
 \hat{\psi}^{\dagger}_{\lambda}(\vec{r}) \hat{\psi}^{\dagger}_{\lambda^{\prime}}(\vecp{r})
 V(\vec{r}-\vecp{r}) 
 \hat{\psi}_{\lambda^{\prime}}(\vecp{r}) \hat{\psi}_{\lambda}(\vec{r}) \label{eqn:Vcontinous}, 
\end{equation} 
where the field operators obey the fermionic anticommutation relations $ \{ \hat{\psi}^{\dagger}_{\lambda}(\vec{r}),\hat{\psi}_{\lambda^{\prime}}(\vecp{r})\} = \delta_{\lambda\lambda^{\prime}}\delta(\vec{r}-\vecp{r})$ and $\lambda$ denotes the spin degree of freedom. In order to place the system on a 3D spatial lattice with lattice spacing $\latconst$ and size $L = \Ns\latconst$ the discretization of the above expression has to be performed.
This leads to the replacement: $\hat{\psi}_{\lambda}(\vec{r})\rightarrow\hat{\psi}_{\lambda}(\vec{r}_{\vec{i}})$, where $\vec{r}_{\vec{i}}=\vec{i}\latconst$ and $\vec{i}=(i_{x},i_{y},i_{z})$ denotes lattice sites, and $i_{x},i_{y},i_{z}=1,\ldots,\Ns$. Consequently one gets instead of (\ref{eqn:Vcontinous}):
\begin{equation}
\hat{V}=\frac{1}{2}\latconst^{6}\sum_{\lambda,\lambda^{\prime}=\uparrow\downarrow}\sum_{\vec{i},\vec{j}}
 \hat{\psi}^{\dagger}_{\lambda}(\vec{r}_{\vec{i}})\hat{\psi}^{\dagger}_{\lambda^{\prime}}(\vec{r}_{\vec{j}})
 V(\vec{r}_{\vec{i}}-\vec{r}_{\vec{j}}) 
 \hat{\psi}_{\lambda^{\prime}}(\vec{r}_{\vec{j}}) \hat{\psi}_{\lambda}(\vec{r}_{\vec{i}}) \label{enq:VDiscrete}.
\end{equation} 
Note that for the problem defined on the lattice, we need only a discrete set of values of the interaction potential defined for the lattice sites $\vec{R}_{k(i,j)}=\vec{r}_{\vec{i}}-\vec{r}_{\vec{j}}$. 
Hence it is sufficient to
introduce the \textit{interaction on the lattice} which is defined 
only on the lattice sites. It can be written in the form:
\begin{equation}
 V(\vec{r}-\vecp{r})=\sum_{k=1}^{D}g_{k}\delta(\vec{r}-\vecp{r}-\vec{R}_{k}). \label{eqn:latticeInteraction}
\end{equation} 
The coupling constants $g_{k}$ contain the full information about
the interaction between the particles. 
The simplest case corresponds to the on-site interaction when $g_{k}$ is nonzero only when 
$\vec{R}_{k}=0$ ($D=1$). A more complicated case when
the neighboring sites are included leads to the so called \textit{extended Hubbard model} ($D=7$), widely used in Monte Carlo simulations (see \cite{BlochEtAl} and references therein). It requires effectively 
two coupling constants (due to the rotational invariance of the Hamiltonian): 
$g_{0}$ for $\vec{R}=(0,0,0)$ (on-site interaction) and $g_{1}$ for $\vec{R}\in\{(\pm\latconst,0,0),(0,\pm\latconst,0),(0,0,\pm\latconst) \}$ (nearest-neighbor interaction). Then the interaction reads:
\begin{equation}
\hat{V}=
 g_{0}\latconst^{3}\sum_{\vec{i}}\dens{\vec{r}_{\vec{i}}}{\uparrow}\dens{\vec{r}_{\vec{i}}}{\downarrow}
 +\frac{g_{1}}{2}\latconst^{3}\sum_{\lambda,\lambda^{\prime}=\uparrow\downarrow}\sum_{<\vec{i},\vec{j}>}
   \dens{\vec{r}_{\vec{i}}}{\lambda}\dens{\vec{r}_{\vec{j}}}{\lambda^{\prime}}, \label{enq:VExtendedHubbard}
\end{equation} 
where $\dens{\vec{r}_{\vec{i}}}{\lambda}=\hat{\psi}^{\dagger}_{\lambda}(\vec{r}_{\vec{i}})\hat{\psi}_{\lambda}(\vec{r}_{\vec{i}})$, and $<\vec{i},\vec{j}>$ denotes the summation over all neighboring pairs. 

\section{$T$ matrix for the lattice interaction}
In order to link the coupling constants $g_{k}$ to the phase shifts one has to consider the $T$ matrix,
which in the case of low energy scattering can be parametrized as:
\begin{equation}
 -\frac{4\pi}{m}T^{-1}_{\vec{p}\vec{p}}\approx-\frac{1}{\sclgth}+\frac{1}{2}\reff p^{2}-ip+O(p^{4})\label{eqn:ERE},
\end{equation}
where $\sclgth$ is the scattering length and $\reff$ denotes the effective range. The $T$ matrix fulfills the Lippmann-Schwinger equation:
\begin{equation}
 T_{\vec{p}\vecp{p}}=V_{\vec{p}\vecp{p}}+
 \frac{1}{L^{3}}\sum_{\vec{k}}V_{\vec{p}\vec{k}}G_{\vecp{p}\vec{k}}T_{\vec{k}\vecp{p}}\label{eqn:Tmatrix},
\end{equation}
where $G_{\vec{p}\vecp{p}}=m/(p^{2}-p^{\prime 2})$ is the free particle propagator with the reduced mass $m/2$. Summation over momenta is limited to the first Brillouin zone ie. \mbox{$-\pi/\latconst\leqslant p_{x,y,z}\leqslant\pi/\latconst$}, and each momentum component is discretized due to 
the box quantization conditions. As a first step, let us rewrite \eqn{eqn:Tmatrix} in an
iterative form, more suitable for numerical applications. Expanding the Lippmann-Schwinger equation one obtains:
\begin{equation}
T_{\vec{p}\vecp{p}}=V_{\vec{p}\vecp{p}}+
\frac{1}{L^{3}}\sum_{\vec{k}}V_{\vec{p}\vec{k}}G_{\vecp{p}\vec{k}}V_{\vec{k}\vecp{p}}+
\frac{1}{L^{6}}\sum_{\vec{k},\vecp{k}}V_{\vec{p}\vec{k}}G_{\vecp{p}\vec{k}}V_{\vec{k}\vecp{k}}G_{\vecp{p}\vecp{k}}V_{\vecp{k}\vecp{p}}+
\ldots\label{eqn:TmatrixIter}
\end{equation}
If we define the matrix $M^{(1)}$ as:
\begin{equation}
M_{\vec{p}\vecp{p}}^{(1)}=\frac{1}{L^{3}}\sum_{\vec{k}}V_{\vec{p}\vec{k}}G_{\vecp{p}\vec{k}}M^{(0)}_{\vec{k}\vecp{p}},\label{eqn:matrixM1}
\end{equation}
where $M_{\vec{p}\vecp{p}}^{(0)}=V_{\vec{p}\vecp{p}}$ then \eqn{eqn:TmatrixIter} takes the form:
\begin{equation}
T_{\vec{p}\vecp{p}}=M_{\vec{p}\vecp{p}}^{(0)}+M_{\vec{p}\vecp{p}}^{(1)}+\frac{1}{L^{3}}\sum_{\vec{k}}V_{\vec{p}\vec{k}}G_{\vecp{p}\vec{k}}M_{\vec{k}\vecp{p}}^{(1)}+\ldots
\end{equation}
Clearly the $T$ matrix can be written as an infinite sum:
\begin{equation}
T_{\vec{p}\vecp{p}}=\sum_{n=0}^{\infty}M_{\vec{p}\vecp{p}}^{(n)},\label{eqn:TmatSeries}
\end{equation} 
where the $M^{(n)}$ matrices are related to each other through the recurrence relation:
\begin{eqnarray}
M_{\vec{p}\vecp{p}}^{(n+1)}&=&\frac{1}{L^{3}}\sum_{\vec{k}}V_{\vec{p}\vec{k}}G_{\vecp{p}\vec{k}}M_{\vec{k}\vecp{p}}^{(n)}\label{eqn:Mm},\\
M_{\vec{p}\vecp{p}}^{(0)}&=&V_{\vec{p}\vecp{p}}.\label{eqn:M0}
\end{eqnarray}
The equation \eqn{eqn:TmatSeries} holds in general case. However in the case when  the effective range expansion \eqn{eqn:ERE} is valid,
only the diagonal matrix elements are required. For the lattice interaction given by \eqn{eqn:latticeInteraction} 
the contribution $M^{(0)}$ takes the following form: 
\begin{equation}
M_{\vec{p}\vec{p}}^{(0)}=\sum_{i=1}^{D}g_{i}e^{-i\vec{p}\rrel_{i}}e^{i\vec{p}\rrel_{i}}=\Tr\,L^{(0)},
\end{equation} 
where the square matrix $L^{(0)}$ of dimension $D\times D$ is defined as:
\begin{equation}
L^{(0)}_{ij}=g_{i}e^{-i\vec{p}(\rrel_{i}-\rrel_{j})}.
\end{equation}
The contribution to $M^{(1)}$ reads:
\begin{eqnarray}
M_{\vec{p}\vec{p}}^{(1)}
&=&\sum_{i,j=1}^{D}g_{i}g_{j}e^{-i\vec{p}(\rrel_{i}-\rrel_{j})}G_{\vec{p}}(\rrel_{j}-\rrel_{i})\nonumber\\
&=&\sum_{i,j=1}^{D}g_{j}L^{(0)}_{ij}G_{\vec{p}}(\rrel_{j}-\rrel_{i})=\Tr\,L^{(1)},%\nonumber\\
%&=&\Tr\,L^{(1)},
\end{eqnarray}
where:
\begin{eqnarray}
L^{(1)}_{ij}&=&\sum_{k=1}^{D}g_{k}L^{(0)}_{ik}G_{\vec{p}}(\rrel_{k}-\rrel_{j}),\\
G_{\vec{p}}(\vec{r})&=&\frac{1}{L^{3}}\sum_{\vec{k}}e^{-i\vec{k}\vec{r}}G_{\vec{p}\vec{k}}.\label{eqn:G_p_r}
\end{eqnarray}
Continuing this procedure for higher order contributions it can be shown that:
\begin{equation}
T_{\vec{p}\vec{p}}=\sum_{n=0}^{\infty}\Tr\,L^{(n)}\label{eqn:T_sum_L},
\end{equation} 
where:
\begin{eqnarray}
L^{(n+1)}_{ij}&=&\sum_{k=1}^{D}g_{k}L^{(n)}_{ik}G_{\vec{p}}(\rrel_{k}-\rrel_{j}),\label{eqn:L_iterative}\\
L^{(0)}_{ij}&=&g_{i}e^{-i\vec{p}(\rrel_{i}-\rrel_{j})}.
\end{eqnarray}
Hence the problem of computing diagonal elements of the $T$ matrix was reduced to calculation of traces of a
relatively small matrices $L^{(n)}$, of dimension $D\times D$. It is important to note that the final result (\ref{eqn:T_sum_L}) includes the effects related to the finite lattice spacing (incorporated by the relative coordinates of the lattice sites $\rrel_{k}$).

Let us examine the large lattice limit, by letting $L\rightarrow\infty$ while keeping the lattice 
constant $\latconst$ fixed. Then the momentum is continuous within 
the first Brillouin zone, and the summation can be replaced by the integration. 
To simplify the analysis, however, we place a spherically symmetric cut-off, including  
only momenta satisfying $p \le \pcut=\pi/\latconst$:
\begin{equation}
 \int d^{3}\vec{p}\rightarrow \int_{0}^{2\pi}dp_{\phi}\int_{0}^{\pi}dp_{\theta}\,\cos p_{\theta}\int_{0}^{\pcut}dp\,p^{2}.
\end{equation}
This prescription sets to zero all two-body matrix elements, if the relative momentum of two particles exceeds a given  momentum cut-off. Note that the function $G_{\vec{p}}(\vec{r})$ will depend on the momentum cut-off and consequently the scattering parameters like $\sclgth$ i $\reff$ 
will also be the functions of $\pcut$.
Performing the integration with respect to the variables $dp_{\phi}$ i $dp_{\theta}$ 
the \eqn{eqn:G_p_r} transforms into:
\begin{equation}
 G_{\vec{p}}(r)=\frac{m}{2\pi^{2}r}\int_{0}^{\pcut}dk\,k\frac{\sin kr}{p^{2}-k^{2}}.
\end{equation} 
The remaining integral can be calculated analytically by expanding  
$\sin kr$ and using the relation:
\begin{equation}
 \frac{1}{x+i0^{+}}=\mathcal{P}\frac{1}{x}-i\pi\delta(x),
\end{equation} 
where $\mathcal{P}$ stands for the principal value. The final result has the form:
\begin{equation}
G_{\vec{p}}(r)=-\frac{m}{4\pi^{2}}\sum_{j=0}^{\infty}\frac{(-1)^{j}r^{2j}}{(2j+1)!}F(j,p)\label{eqn:G_integrated}
\end{equation}
where: 
\begin{equation}
 F(j,p)=\sum_{l=0}^{j}\frac{2p^{2l}\pcut^{2j-2l+1}}{2j-2l+1} + p^{2j+1}\ln\vert\frac{\pcut-p}{\pcut+p}\vert + i\pi p^{2j+1}.
\end{equation}
The expression (\ref{eqn:T_sum_L}) together with (\ref{eqn:G_integrated}) and (\ref{eqn:ERE}) provides the most convenient prescription for numerical applications.

\section{Example: on-site interaction}
As an example let us consider the on-site interaction:
\begin{equation}
V(\vec{r}-\vecp{r})=g_{0}\delta(\vec{r}-\vecp{r}).
\end{equation} 
It corresponds to the well-known \textit{Hubbard model} and is presently widely used to simulate the system
of dilute, cold fermionic atoms.\cite{bcs-bec,BurovskiEtAl,LeeSchaefer} In this particular case, it is possible to find an analytic formula relating the coupling constant $g_{0}$ to the scattering length and the effective range. It is easy to realize that $L^{(n)}$ become now one-dimensional matrices ie.:
\begin{equation}
 L^{(n)}=g_{0}\,(g_{0}G_{\vec{p}})^{n},
\end{equation} 
where:
\begin{equation}
 G_{\vec{p}}=G_{\vec{p}}(0)=-\frac{m}{4\pi^{2}}\left( 2\pcut-p\,\ln\vert\frac{p+\pcut}{p-\pcut}\vert+i\pi p\right) .
\end{equation} 
The sum of the geometric series (\ref{eqn:T_sum_L}) reads:
\begin{equation}
 T_{\vec{p}\vec{p}}=\frac{g_{0}}{1-g_{0}G_{\vec{p}}}.
\end{equation}
In order to find the inverse of the diagonal matrix elements of $T$ for small values of the momentum,
we expand the logarithmic function:
\begin{equation}
 \ln\vert\frac{p+\pcut}{p-\pcut}\vert\approx\frac{2}{\pcut}p+\frac{2}{3\pcut^{3}}p^{3},
\end{equation} 
and get:
\begin{equation}
-\frac{4\pi}{m}T^{-1}_{\vec{p}\vec{p}}= -\frac{4\pi}{mg_{0}} -  \frac{2\pcut}{\pi}+\frac{2}{\pi\pcut}p^{2}-ip+O(p^{4}).
\end{equation} 
Comparing this equation with \eqn{eqn:ERE} one reproduces the well-known results\cite{bcs-bec}:
\begin{equation}
\frac{1}{\sclgth}=\frac{4\pi}{mg_{0}}+\frac{2\pcut}{\pi},\quad\reff=\frac{4}{\pi\pcut}.
\end{equation}
It is worth noting that due to the finite lattice spacing the zero range potential acquires 
the non-zero effective range $\reff$.

\section{Conclusions}
In this work we have presented the method to determine the two-body effective interaction parameters from 
the low energy scattering data when the problem is defined on the lattice. In the limit of 
large lattices we have obtained expressions which are particularly convenient for numerical applications. 
The prescription relates the interaction coupling constants to the effective range expansion parameters 
and can be applied to Monte Carlo simulations of many-body systems on large lattices.

\section*{Acknowledgments}
This work has been partially supported by the Polish Ministry of Science 
under contracts No. N N202 328234, N N202 110236 and by the UNEDF SciDAC Collaboration 
under DOE grant DE-FC02-07ER41457. One of the authors (G.W.)
acknowledges the support within 
The Integrated Regional Operational Programme: ``Mazowieckie Stypendium Doktoranckie''.

\end{document}